\documentclass[prl,twocolumn,aps]{revtex4-1}
\usepackage{graphicx}
\usepackage{amsfonts}
\usepackage{hyperref}
\begin{document}
\newcommand{\bra}{\langle}
\newcommand{\ket}{\rangle}
\newcommand{\al}{\alpha}
\newcommand{\be}{\beta}
\newcommand{\ga}{\gamma}
\newcommand{\de}{\delta}
\newcommand{\D}{\Delta}
\newcommand{\ep}{\epsilon}
\newcommand{\varep}{\varepsilon}
\newcommand{\e}{\eta}
\renewcommand{\th}{\theta}
\newcommand{\Th}{\Theta}
\newcommand{\la}{\lambda}
\newcommand{\La}{\Lambda}
\newcommand{\Ga}{\Gamma}
\newcommand{\m}{\mu}
\newcommand{\n}{\nu}
\renewcommand{\r}{\rho}
\newcommand{\si}{\sigma}
\newcommand{\Si}{\Sigma}
\newcommand{\ta}{\tau}
\newcommand{\vp}{\varphi}
\newcommand{\p}{\phi}
\renewcommand{\c}{\chi}
\newcommand{\ps}{\psi}
\renewcommand{\o}{\omega}
\renewcommand{\O}{\Omega}
\newcommand{\OO}{{\cal O}}
\newcommand{\C}{{\cal C}}
\newcommand{\pa}{\partial}
\newcommand{\beq}{\begin{equation}}
\newcommand{\eeq}{\end{equation}}
\newcommand{\mc}{\mathcal}
\newcommand{\mb}{\mathbb}
\newcommand{\mycomment}[1]{}
\newcommand{\wh}[1]{\widehat{#1}}
\newcommand{\fk}{f_\textup{bulk}}
\newcommand{\fy}{f_\textup{bdy}}
\newcommand{\yi}{y_i}
\newcommand{\yl}{y_L}
\newcommand{\yr}{y_R}

\newcommand{\nn}{\nonumber}
\newcommand{\Sl}{\sum\limits}
\newcommand{\blue}{\color{blue}}
\newcommand{\red}{\color{red}}
\newcommand{\green}{\color{green}}
\newcommand{\black}{\color{black}}

\def\msout#1{\textrm{\sout{#1}}}
\def\mxout#1{\textrm{\xout{#1}}}

 \def\be{\begin{equation}}
\def\ee{\end{equation}}
\def\bea{\begin{eqnarray}}
\def\eea{\end{eqnarray}}
\def\nn{\nonumber}

\def\tr{{\mbox{tr}}}
\def\Atr{{\mbox{Tr}}}

\def\a{\alpha}
\def\b{\beta}
\def\g{\gamma}
\def\d{\delta}
\def\lam{\lambda}
\def\u{\mu}
\def\v{\nu}
\def\r{\rho}
\def\t{\tau}
\def\z{\zeta}
\def\s{\sigma}
\def\th{\theta}

\def\te{\tilde{e}}
\def\tK{\tilde{K}}
\def\tB{\tilde{B}}
\def\htK{\hat{\tilde{K}}}

\def\Qh{\hat{Q}}
\def\baret{\overline{\eta}}
\def\homega{{\hat{\omega}}}
\def\bpsi{{\overline{\psi}}}
\def\wtau{{\widetilde{\tau}}}
\def\bth{{\overline{\theta}}}
\def\blam{{\overline{\lambda}}}

\def\da{{\dot{\a}}}
\def\db{{\dot{\b}}}
\def\dg{{\dot{\g}}}

\def\bj{{\overline{j}}}
\def\bk{{\overline{k}}}
\def\bz{{\overline{z}}}
\def\sa{{\hat{a}}}
\def\sb{{\hat{b}}}
\def\sc{{\hat{c}}}
\def\wa{{\tilde{a}}}
\def\wb{{\tilde{b}}}
\def\wc{{\tilde{c}}}
\def\oa{{\overline{a}}}
\def\ob{{\overline{b}}}
\def\oc{{\overline{c}}}
\def\od{{\overline{d}}}

 \def\CA{{\cal A}}
 \def\CC{{\cal C}}
 \def\CF{{\cal F}}
 \def\CI{{\cal I}}
 \def\cJ{{\cal J}}
 \def\tJ{{\tilde J}}
 \def\tcJ{{\tilde\cal J}} 
 \def\CO{{\cal O}}
 \def\o{{\rm ord}}
 \def\Ph{{\Phi }}
 \def\L{{\Lambda}}
 \def\CN{{\cal N}}
 \def\p{\partial}
 \def\pslash{\p \llap{/}}
 \def\Dslash{D \llap{/}}
 \def\apm{{\a^{\prime}}}
 \def\r{\rightarrow}
\def\ts{\tilde s}
\def\tu{\tilde u}
 \def\BR{\IR}
 \def\BZ{\IZ}
 \def\BC{\IC}
 \def\BM{\QM}
 \def\BP{\IP}
 \def\BH{\QH}
 \def\BX{\QX}
 \def\sym#1{{{\rm SYM}} _{#1 +1}}
 \def\imp{$\Rightarrow$}
 \def\IZ{\relax\ifmmode\mathchoice
 {\hbox{\cmss Z\kern-.4em Z}}{\hbox{\cmss Z\kern-.4em Z}}
 {\lower.9pt\hbox{\cmsss Z\kern-.4em Z}}
 {\lower1.2pt\hbox{\cmsss Z\kern-.4em Z}}\else{\cmss Z\kern-.4em Z}\fi}
 \def\IB{\relax{\rm I\kern-.18em B}}
 \def\IC{{\relax\hbox{$\inbar\kern-.3em{\rm C}$}}}
 \def\Ic{{\relax\hbox{$\inbar\kern-.22em{\rm c}$}}}
 \def\ID{\relax{\rm I\kern-.18em D}}
 \def\IE{\relax{\rm I\kern-.18em E}}
 \def\IF{\relax{\rm I\kern-.18em F}}
 \def\IG{\relax\hbox{$\inbar\kern-.3em{\rm G}$}}
 \def\IGa{\relax\hbox{${\rm I}\kern-.18em\Gamma$}}
 \def\IH{\relax{\rm I\kern-.18em H}}
 \def\II{\relax{\rm I\kern-.18em I}}
 \def\IK{\relax{\rm I\kern-.18em K}}
 \def\IP{\relax{\rm I\kern-.18em P}}

\def\Tr{{\rm Tr}}
 \font\cmss=cmss10 \font\cmsss=cmss10 at 7pt
 \def\IR{\relax{\rm I\kern-.18em R}}

\def\wdg{{\wedge}}

\newcommand\ev[1]{{\langle {#1}\rangle}}
\newcommand\SUSY[1]{{{\cal N} = {#1}}}
\newcommand\diag[1]{{\mbox{diag}({#1})}}
\newcommand\com[2]{{\left\lbrack {#1}, {#2}\right\rbrack}}

\newcommand\px[1]{{\partial_{#1}}}
\newcommand\qx[1]{{\partial^{#1}}}

\newcommand\rep[1]{{\bf {#1}}}

\def\gam{{\widetilde{\gamma}}} 
\def\sig{{\sigma}} 
\def\hsig{{\hat{\sigma}}} 

\def\eps{{\epsilon}} 

\def\bZ{{\overline{Z}}}
\def\BR{{\mathbb R}}
\def\Lag{{\cal L}}
\def\cO{{\cal O}}
\def\cH{{\cal H}}
\def\wcO{{\widetilde{\cal O}}}
\def\vL{{\vec{L}}}
\def\vx{{\vec{x}}}
\def\vy{{\vec{y}}}

\def\vLf{{\vec{\lambda}}}

\newcommand\cvL[1]{{L^{#1}}} 

\def\npsi{{\psi^{(inv)}}}
\def\bnpsi{{\bpsi^{(inv)}}}
\def\wpsi{{\widetilde{\psi}}}
\def\bwpsi{{\overline{\widetilde{\psi}}}}
\def\hpsi{{\hat{\psi}}}
\def\bhpsi{{\overline{\hat{\psi}}}}

\newcommand\nDF[1]{{{D^F}_{#1}}}
\def\wA{{\widetilde{A}}}
\def\wF{{\widetilde{F}}}
\def\wJ{{\widetilde{J}}}
\def\hA{{\hat{A}}}
\def\hF{{\hat{F}}}
\def\hJ{{\hat{J}}}
\def\MapL{{\Upsilon}}
\def\hR{{\hat{R}}}
\def\hL{{\hat{L}}}

\def\cpl{{\lambda}}  
\def\mcr{{\mathcal{R}}}
\def\xv{{\vec{x}}}
\def\xvt{{\vec{x}^{\top}}}
\def\nv{{\hat{n}}}
\def\nvt{{\hat{n}^{\top}}}
\def\hM{{M}}
\def\hgM{{\widetilde{M}}}
\def\utr{{\mbox{tr}}}

\newcommand{\fig}[1]{fig.\ (\ref{#1})}
\def\dd{\mbox{d}}
\def\ddd{\mbox{\sm d}}
\def\o{\omega}
\def\bra{\langle}
\def\ket{\rangle}
\def\a{\alpha}
\def\b{\beta}
\def\d{\delta}
\def\dd{\partial}
\def\D{\Delta}
\def\LL{\triangle}
\def\g{\gamma}
\def\G{\Gamma}
\def\e{\epsilon}
\def\ve{\varepsilon}
\def\et{\eta}
\def\f{\phi}
\def\F{\Phi}
\def\vf{\varphi}
\def\k{\kappa}
\def\l{\lambda}
\def\L{\Lambda}
\def\m{\mu}
\def\n{\nu}
\def\s{\sigma}
\def\S{\Sigma}
\def\o{\omega}
\def\p{\pi}
\def\r{\rho}
\def\t{\tau}
\def\th{\theta}
\def\vt{\vartheta}
\def\ra{\rightarrow}
\def\la{\leftarrow}
\def\pa{\partial}
\def\ov{\overline}
\def\Pl{s_{\sm{Pl}}}
\def\tr{\tilde R}
\def\td{\tilde d}
\def\gmn{g_{\mu \nu}}
\def\DO{\D_2}
\def\O{{\cal O}}

\newcommand{\ti}[1]{\tilde{#1}}
\renewcommand{\^}[1]{\hat{#1}}
\newcommand{\sm}[1]{\mbox{\scriptsize #1}}
\newcommand{\tn}[1]{\mbox{\tiny #1}}
\renewcommand{\@}[1]{\sqrt{#1}}
\renewcommand{\le}[1]{\label{#1}\end{eqnarray}}
\newcommand{\eq}[1]{(\ref{#1})}
\def\nn{\nonumber\\}
\def\nm{\nonumber}
\newcommand{\rf}[1]{\cite{ref:#1}}
\newcommand{\rr}[1]{\bibitem{ref:#1}}
\def\qu{\ {\buildrel{\displaystyle ?} \over =}\ }
\def\smqu{\ {\buildrel ?\over =}\ }
\def\ffract#1#2{\raise .35 em\hbox{$\scriptstyle#1$}\kern-.25em/
\kern-.2em\lower .22 em \hbox{$\scriptstyle#2$}}
\def\GN{G_{\mbox{\tn N}}}
\def\lPl{s_{\mbox{\tn{Pl}}}}
\def\fn{f_{(0)}}
\def\fe{f_{(1)}}
\def\ft{f_{(2)}}
\def\fd{f_{(3)}}
\def\Ric{{\mbox{Ric}}}
\def\Rie{{\mbox{Rie}}}
\def\Ein{{\mbox{Ein}}}
\def\nl{\newline}
\def\cl{\textcolor}
\def\clb{\colorbox}
\def\na{\nabla}
\def\half{{1\over2}\,}
\def\nonu{\nonumber \\{}}
\def\da{\dot{a}}
\def\db{\dot{b}}
\def\dc{\dot{c}}
\def\tg{\tilde{\g}}
\def\bF{\bar{F}}


\def\vwsX{{\bf X}} 
\def\dvwsX{{\dot{\bf X}}} 
\newcommand\vxmod[1]{{{\bf X}_{#1}}} 
\newcommand\dvxmod[1]{{\dot{{\bf X}}_{#1}}} 

\def\vwsP{{\bf \Pi}} 

\def\wsX{{X}} 
\def\wsZ{{Z}} 
\def\bwsZ{{\overline{Z}}} 
\def\wsPsi{{\Psi}} 
\def\bwsPsi{{\overline{\Psi}}} 

\def\twsZ{{\widetilde{Z}}} 
\def\twS{{\widetilde{S}}} 
\def\btwsZ{{\overline{\widetilde{Z}}}} 
\def\twsPsi{{\widetilde{\Psi}}} 
\def\btwsPsi{{\overline{\widetilde{\Psi}}}} 
\def\tV{{\widetilde{V}}}
\def\talp{{{\tilde{\a}}'}} 

\def\hLf{{\hat{\lambda}}}
\def\hL{{\hat{L}}}


\newdimen\tableauside\tableauside=1.0ex
\newdimen\tableaurule\tableaurule=0.4pt
\newdimen\tableaustep
\def\phantomhrule#1{\hbox{\vbox to0pt{\hrule height\tableaurule width#1\vss}}}
\def\phantomvrule#1{\vbox{\hbox to0pt{\vrule width\tableaurule height#1\hss}}}
\def\sqr{\vbox{%
  \phantomhrule\tableaustep
  \hbox{\phantomvrule\tableaustep\kern\tableaustep\phantomvrule\tableaustep}%
  \hbox{\vbox{\phantomhrule\tableauside}\kern-\tableaurule}}}
\def\squares#1{\hbox{\count0=#1\noindent\loop\sqr
  \advance\count0 by-1 \ifnum\count0>0\repeat}}
\def\tableau#1{\vcenter{\offinterlineskip
  \tableaustep=\tableauside\advance\tableaustep by-\tableaurule
  \kern\normallineskip\hbox
    {\kern\normallineskip\vbox
      {\gettableau#1 0 }%
     \kern\normallineskip\kern\tableaurule}%
  \kern\normallineskip\kern\tableaurule}}
\def\gettableau#1 {\ifnum#1=0\let\next=\null\else
  \squares{#1}\let\next=\gettableau\fi\next}

\tableauside=1.0ex
\tableaurule=0.4pt



\newcommand{\eeqn}{\end{eqnarray}}
\newcommand{\ack}[1]{{\bf Pfft! #1}}
\newcommand{\osigma}{\overline{\sigma}}
\newcommand{\orho}{\overline{\rho}}
\newcommand{\myfig}[3]{
	\begin{figure}[ht]
	\centering
	\includegraphics[width=#2cm]{#1}\caption{#3}\label{fig:#1}
	\end{figure}
	}
\newcommand{\littlefig}[2]{
	\includegraphics[width=#2cm]{#1}}

\title{
Generalized Wilson-Fisher critical points from the conformal OPE}
\author{{\bf Ferdinando~Gliozzi$^a$, Andrea~L.~Guerrieri$^{b,f}$, Anastasios~C.~Petkou$^{c}$ and Congkao~Wen$^{d,e,f}$ }}
\affiliation{$^a$ Dipartimento di Fisica, Università di Torino
and Istituto Nazionale di Fisica Nucleare - sezione di Torino
Via P. Giuria 1 I-10125 Torino, Italy.\\
  $^{b}$ Department of Physics, Faculty of Science, Chulalongkorn University,
Thanon Phayathai, Pathumwan, Bangkok 10330, Thailand. \\
$^{c}$ Institute of Theoretical Physics, Aristotle University of Thessaloniki, Thessaloniki 54124, Greece.
\\ $^{d}$ Walter Burke Institute for Theoretical Physics, California Institute of Technology, Pasadena, CA  91125
\\ $^{e}$ Mani L. Bhaumik Institute for Theoretical Physics, Department of Physics and Astronomy, UCLA, Los Angeles, CA 90095
\\ $^{f}$ I.N.F.N. Sezione di Roma Tor Vergata, Via della Ricerca Scientifica
00133 Roma, Italy
}


\date{\today}
\begin{abstract}
We study possible smooth deformations of Generalized Free Conformal 
Field Theories in arbitrary dimensions by exploiting the singularity structure 
of the conformal blocks dictated by the null states. We derive in this way, at 
the first non trivial order in the $\epsilon$-expansion, the anomalous 
dimensions of an infinite class of scalar local operators,  without using 
the equations of motion. In the cases where other computational 
methods apply, the results agree.
\end{abstract}

\maketitle
\section{Introduction}

The remarkable success of the numerical conformal bootstrap 
\cite{ElShowk:2012ht,El-Showk:2014dwa,Kos:2014bka,Kos:2016ysd} calls for an analytical explanation of the {\sl  unreasonable effectiveness} of conformal field theory 
(CFT). It is therefore pertinent to ask whether CFT techniques can reproduce, and eventually surpass in accuracy, the perturbative results for critical indices that have been accumulated over the years for a variety of fixed point theories in different dimensions. This question has been recently asked by \cite{Rychkov:2015naa} in the context of the $\phi^4$ theory in $d=4-\epsilon$ dimensions. It was shown there that the critical exponents can be reproduced under the following three assumptions: I) The perturbative Wilson-Fisher (WF)  fixed point is described by a CFT. II) In 
the $\epsilon\to0$ limit correlation functions approach those of the free theory. III) The equations of motion describe the transformation of  a primary operator in the free theory into a descendant at the WF fixed point.  Such an approach has been generalized more recently in \cite{Basu:2015gpa, Nii:2016lpa,Hasegawa:2016piv}, see also \cite{Gopakumar:2016wkt, Gopakumar:2016cpb}.

Motivated by the same questions we aim to extend the above ideas to the vast class of generalized free CFTs (GFCFTs) in arbitrary dimensions in order to study their nearby WF fixed points. We show that requiring II) above makes assumption III) redundant since the transformation of free primary operators into descendants of the interacting theory is already dictated by the analytic structure of  the null states of the GFCFTs. Then, without the use of equations of motion we calculate the leading order critical quantities in a variety of models in diverse dimensions $d$. For the known cases our results agree with previous calculations.

The 4pt  function of arbitrary scalar operators in a generic $d$-dimensional CFT
can be parametrised as
\bea
&& \bra \OO_1(x_1)\OO_2(x_2)\OO_3(x_3)\OO_4(x_4)\ket=
\cr
&& 
\frac{g(u,v)}
{\vert x_{12}\vert^{\Delta^{+}_{12}}
\vert x_{34}\vert^{\Delta^{+}_{34}}} \left( {\vert x_{24}\vert \over \vert x_{14}\vert} \right)^{\Delta^{-}_{12}}
\left( {\vert x_{14}\vert \over \vert x_{13}\vert} \right)^{\Delta^{-}_{34}} \,,
\label{4pt}
\eea
where $\Delta^{\pm}_{ij} = \Delta_{i} \pm \Delta_{j}$ and $\Delta_{i}$ is the scaling dimension of $\OO_i$, while $u={x_{12}^2 x_{34}^2 \over x_{13}^2 x_{24}^2}$ and $v={x_{14}^2 x_{23}^2 \over x_{13}^2 x_{24}^2}$ are the cross ratios. This corresponds to inserting the conformal OPE in the direct channel, in which case the function $g(u,v)$ can be expanded in terms of conformal blocks
$G^{a,b}_{\Delta,\ell}(u,v)$, i.e. eigenfunctions of the quadratic 
Casimir operator of SO$(d+1,1)$:
\be
g(u,v)=\sum_{\Delta,\ell}c_{\Delta,\ell} G^{a,b}_{\Delta,\ell}(u,v).
\ee
Here $a=-\Delta^-_{12}/2$ and $b=\Delta^-_{34}/2$, and 
$\Delta$ and $\ell$ are the scaling dimensions and the spin of the 
primary operators that are exchanged. 
When $\Delta$ takes some particular values the generic conformal blocks are singular. It has been shown in \cite{Kos:2014bka} that the singularities are poles and they are associated to null states as \beq
G^{a,b}_{\Delta,\ell}=F^{a,b}_{\Delta,\ell}+\sum_k R^{a,b}_k\frac{G^{a,b}_{\Delta'_k,\ell'_k}}
{\Delta-\Delta_k}\,,
\label{generalcb}
\eeq
where $F$ is an entire function of $\Delta$. The poles in (\ref{generalcb}) reflect the 
contribution 
of a null state at  $[\Delta_k,\ell_k]$; the form of the residue tells 
us that if 
$R^{a,b}_k\not=0$ such a null state has a descendant at $[\Delta_k',\ell_k']$.  
The explicit formulas for $[\Delta_k,\ell_k]$, $R^{a,b}_k$ and  
 $[\Delta_k',\ell_k']$ can be found in  \cite{Kos:2014bka}.
Another, equivalent, way to obtain them is to consider the following 
expansion,
valid  for arbitrary  $a,b,d $ and $\delta$,
\beq
u^\delta=\sum_{\tau=0}^\infty\sum_{\ell=0}^\infty 
\frac{(-1)^\ell(2\nu)_\ell}{\tau!\ell!(\nu)_\ell(\nu+\ell+1)_\tau}
c^{a,b}_{\delta,\tau,\ell}
G^{a,b}_{2\delta+2\tau+\ell, \ell}(u,v),
\label{ud}
\eeq 
here $(x)_y=\frac{\Gamma(x+y)}{\Gamma(x)}$ is the Pochhammer symbol
and $\nu=\frac d2 -1$. The coefficients above are given by
\beq
\label{cab}
c^{a,b}_{\delta,\tau,\ell}=\frac{\prod_{i=a,b}(i+\delta)_{\ell +\tau}
(i+\delta-\nu)_\tau}{
(\Delta-1)_\ell(\Delta-\nu-\tau-1)_\tau(\Delta-2\nu-\tau-\ell-1)_\tau}
\eeq
with $\Delta=2\delta+2\tau+\ell$. The three families of simple poles of 
$ c^{a,b}_{\delta,\tau,\ell}$  correspond exactly to the null 
states
of  (\ref{generalcb}). Requiring the cancellation of these 
poles in the RHS of (\ref{ud}) yields the explicit form of the residues 
of (\ref{generalcb})  \cite{longpaper} which precisely produce all the results of \cite{Kos:2014bka}. 
In the following we need only the dimensions and the residues of those 
scalar null states  having a scalar descendant. Their scaling dimensions 
$\Delta_k$ and those of the scalar descendants $\Delta_k'$ are labelled by an integer $k$:
\beq
\Delta_k=\frac d2-k\,,~\Delta_k'=\frac d2+k\,,\;\;k=1, 2 \, , \dots \, ,
\label{null}
\eeq
and the corresponding residue is given by
\beq
R^{a,b}_k=-\frac{(-1)^k k \prod_{c=\pm a,\pm b}(\frac d4-\frac k2-c)_k}
{(k!)^2(\frac d2-k-1)_{2k}}\,.
\label{residue}
\eeq
Notice that $\Delta_k+\Delta_k'=d$ which means the above operators are {\it shadows} of each other. 

We want to apply such universal properties of the conformal blocks to the 
study of smooth deformations of GFCFTs. 
Scalar GFCFTs are constructed by a single elementary scalar field $\phi_f$ 
with dimension $\Delta_{\phi_f}$. Its two-point function is given by
$\langle\phi_f(x_1)\phi_f(x_2)\rangle =1/x_{12}^{2\Delta_{\phi_f}}$
and the three-point function of $\phi_f$ vanishes. All other correlation functions, 
either of $\phi_f$ or of its composites, are given by Wick contractions.
We are particularly interested to the cases in which $\Delta_{\phi_f}=\frac d2 -k$. They correspond to GFCFT  the have a Lagrangian description as massless free theories with $\partial^{2k}$ kinetic term that can be coupled to the stress energy tensor \cite{Diab:2016spb, Guerrieri:2016whh, Osborn:2016bev, Brust:2016gjy}.
The  $k=1$ case corresponds to the free canonical theories. 
Some GFCFTs with $k>1$ play an important role in the study of $1/N$ 
expansions in Gross-Neveu and $O(N)$ vector models in high dimensions 
\cite{Diab:2016spb, Guerrieri:2016whh, Osborn:2016bev}. Applying the Wick 
contractions to the product $\phi_f^p(x)\phi_f^{p+1}(y)$ 
generates the OPE (see Eq. (\ref{pp1})) 
\be
[\phi^p_f] \times [\phi^{p+1}_f]=\sqrt{p+1}[\phi_f]+\dots \, .
\label{ope}
\ee
Throughout the paper, we normalize the operator as $[\phi^p_f] = \phi^p_f/\sqrt{p!}$ such that the two-point function takes the following simple form, 
\bea
\langle [\phi^p_f](x_1) \,\, [\phi^p_f](x_2)\rangle = {1 \over (x^2_{12})^{\Delta_{\phi^p_f}} } \, ,
\eea
where $\Delta_{\phi^p_f}$, the dimension of the composite operator $[\phi^p_f]$, is simply $p\,\Delta_{\phi_f}$ for a free theory. 
Since $\Delta_{\phi_f}=\frac d2-k$, there is  a possible contribution of the 
null state (\ref{null}), however, according to (\ref{residue}), 
the residue $R^{a,b}_k$ is zero. We will deform this free theory by 
replacing $\phi_f^p\to\phi^p$ with
\be \label{eq:anomalousd}
\Delta_{\phi^p}=\Delta_{\phi^p_f}+\gamma_p=\Delta_{\phi^p_f}+\gamma^{(1)}_p\epsilon+\gamma^{(2)}_p\epsilon^2+\dots \, .
\ee
We also deform the relevant OPE coefficients such that they have a smooth $\epsilon\rightarrow 0 $ limit to their corresponding free values.
As a consequence now $R^{a,b}_k\propto \epsilon^2\not=0$ and in the conformal block 
$G_{\Delta_\phi,0}^{a,b}$ a new sub-representation associated to a 
descendant of dimension $\Delta_k'=\frac d2+k$ appears. 
If in the free field theory OPE there is a primary operator of dimension 
$\Delta_k'$ we say that the theory is smoothly deformable. In the interacting 
theory such  an operator is promoted to a descendant of the null state. The 
matching of $R^{a,b}_k$ with the OPE coefficient of the free theory yields an 
equation for the anomalous dimensions of the involved operators. 
The systematic study of these equations allows us to find the first 
non-trivial anomalous dimensions of $\phi^p$ for a huge class of GFCFTs, in arbitrary dimensions. Our results include the known cases of Wilson-Fisher points near 
canonical free field theories. Equations (\ref{eq:finding1}, \ref{eq:finding2}), and (\ref{eq:finding3}, \ref{eq:finding4}, \ref{eq:finding5}, \ref{eq:finding6}, \ref{eq:finding7}) show our main findings. 
\section{The Generalized Wilson-Fisher critical points}
Our  general strategy to find smooth deformations of GFCFTs consists in 
constructing
a suitable 4pt function of a generalized free theory at arbitrary space 
dimension $d$ and switching on interaction  by simply  replacing 
$d\to d-\epsilon$ and $\Delta_{\phi^p_f}\to \Delta_{\phi^p}$ according to 
(\ref{eq:anomalousd}). As discussed in the previous section this procedure 
defines actually a smooth deformation of the nearby free OPE only if does not  
generate new primary operators in the limit $\epsilon\to 0$.  
In this way we are essentially studying Generalized Wilson-Fisher (GWF) 
critical points. Given that we start with arbitrary $\Delta$  and $d$ 
we have a huge theory space to explore. In the following we will discuss 
just  a few nontrivial examples that include most of the known results 
obtained for the Wilson-Fisher critical points 
by other methods, but also yield some new results.

\subsection{The $\phi^{2n}$ critical points}

Consider the generic free OPE
\be
\label{pp1}
[\phi^p_f] \times [\phi^{p+1}_f]=\sum_{n=1}^{p+1} \lambda_{p,p,2n-1}
[\phi^{2n-1}_f]+\text{spinning blocks} \, ,
\ee
with the OPE coefficients being
\be
\lambda_{p,p,2n-1}=B_{2n-1,n}\sqrt{\frac{p+1}{(2n-1)!}}(p-n+2)_{n-1}\,,
\ee
here $B_{n,m}$ is the binomial coefficient. For $\Delta_{\phi_f}=d/2-k,\,k=1,2,3, \ldots$, inserting (\ref{pp1}) into the direct channel of the 4pt function $\bra\phi^p\phi^{p+1}\phi^p\phi^{p+1}\ket$  one obtains an expansion in terms, among others, of scalar conformal blocks of the type 
$G^{-a_f,a_f}_{\Delta_{\phi^{2n-1}_f}}(u,v)$ with $a_f=\Delta_{\phi_f}/2$. When we smoothly deform the theory, each OPE coefficient should be modified with a term which vanishes in the limit of $\epsilon \rightarrow 0$. Most importantly the operator $\phi^{2n-1}$ becomes a descendant, which should be removed from the OPE. As we discussed in the previous section, the conformal block in the interacting theory $G^{-a,a}_{\Delta_\phi}$ has a singularity with a residue proportional to the conformal block $G^{-a,a}_{{d \over 2}+k}$ which is precisely the missing conformal block of the operator $\phi^{2n-1}$ in the free theory in the limit $\epsilon \rightarrow 0$. For the interacting theory, we have $a=a_f + \gamma_{{p+1}}-\gamma_{p}$ and $\Delta_\phi=\Delta_{\phi_f} + \gamma_{1}$, with $\gamma_n$ being the anomalous dimensions of $\phi^n$ as defined in (\ref{eq:anomalousd}). Matching the operator dimensions requires $d=2nk/(n-1)$. 
Using then the explicit result for the residue in (\ref{residue}), the matching condition gives
\be
\lambda^2_{p,p,1} {R_k^{-a,a} \over \Delta_\phi - \Delta_{\phi_f}} =\lambda^2_{p,p,2n-1} \, .
\ee
Using explicit form of $R_k^{-a,a}$ in (\ref{residue}) and performing the $\epsilon$-expansion yields
\be
(p+1)\frac{(-1)^{k+1}\left( \frac{d}{2} -k\right)_k}{4k\left( \frac{d}{2} \right)_k}\frac{(\gamma_{{p+1}}^{(1)}-\gamma_{p}^{(1)})^2 \epsilon^2 + \ldots }{\gamma_{1}^{(1)} \epsilon + \gamma_{1}^{(2)} \epsilon^2  + \ldots }=\lambda^2_{p,p,2n-1} \, .
\ee 
Since the RHS is finite, this immediately implies $\gamma_{1}^{(1)}=0$.  
The above matching equation then leads to the following recursion relation
\be
\gamma_{{p+1}}^{(1)}-\gamma_{p}^{(1)}= \kappa(k,n) (p-n+2)_{n-1}\, ,
\ee
with 
\be \label{eq:kappa}
\kappa^2(k,n)=\frac{(-1)^{k+1}4 k \,B^2_{2n-1,n}\left(\frac{n k}{n-1} \right)_k}{(2n-1)!\left(\frac{k}{n-1}\right)_k}\gamma_1^{(2)} \,,
\ee
where we used $d=2nk/(n-1)$. The apparent sign ambiguity of $\kappa(k,n)$ will be solved in a moment. Using the fact $\gamma_{0}=0$ the recursion 
relation gives
\be \label{eq:solution}
\gamma_{p}^{(1)}= \frac{\kappa(k,n)}{n}(p-n+1)_n.
\ee
The crucial observation is that the operator $\phi^{2n-1}$ that becomes a descendant of $\phi$ has a fixed dimension, namely ${d \over 2} + k$. From this fact we deduce that its anomalous dimension is $\gamma^{(1)}_{{2n-1}}=n-1$. Using this we fix $\kappa(k,n) = n(n-1)/(n)_n$, and plugging this back into (\ref{eq:solution}) yields
\bea \label{eq:finding1}
\gamma_{p}^{(1)}= \frac{(n-1)}{(n)_n}(p-n+1)_n \, ,
\eea
which is interestingly independent of $k$. 
From (\ref{eq:kappa}), we then obtain the anomalous dimension of $\phi$ at order $\epsilon^2$
\bea \label{eq:finding2}
\gamma^{(2)}_1 = (-1)^{k+1} 2 { n \left( {k \over n-1} \right)_k \over k \left( {n \, k \over n-1} \right)_k } (n-1)^2 \left[ { (n!)^2 \over (2n)!} \right]^3 \, .
\eea
A few comments are in order: in the case of $k=1$, namely for the canonical scalar, $\gamma^{(2)}_1$ reduces to  $2(n{-}1)^2 \left[ { (n!)^2 \over (2n)!} \right]^3$ which is well-known multicritical result \cite{book}: for $n=2$ it corresponds to the  $\phi^4$ theory in $d=4-\epsilon$ while $n=3$  to the $\phi^6$ theory in $d=3-\epsilon$. More generally when $k>1$, it is a smooth deformation of scalar GFCFT with $\Delta_{\phi}={d \over 2}-k={k \over n-1}$ in $d={2nk \over n-1}$. For $k>1$, we have assumed that we only turn on one possible marginal deformation of the form $\phi^{2n}$, in principle one may have marginal interactions with derivatives. Notice also that $k>1$ allows us to study multicritical, but non unitary, theories in integer dimensions $d>6$.

\subsection{O$(N)$ models}
Here we apply our method to theories with global O$(N)$ symmetry. We consider scalars $\phi_i$, $i=1,2,..N$ with $\Delta_{\phi_i} = d/2-k$ in $d=3k$ and $d=4k$ as our examples. We denote $\sum_i \phi_i\phi_i = \phi^2$ and consider the following free OPEs
\bea
\!\!\!&&  [\phi_i (\phi^2)^{p-1}] \times [(\phi^2)^p] = \sqrt{2p \over N} \left( [\phi_i]  
+ \sqrt{ (6p{+}N{-}4)^2 \over 2(N+2)}   [ \phi_i \phi^2 ] \right.
\cr
\!\!\! &+&  \left. \sqrt{2 (10p+3N-8)^2 (p-1)^2\over  (N+4)(N+2)} [ \phi_i (\phi^2)^2 ] \right) + \ldots \, , \cr
\!\!\!&&[(\phi^2)^p] \times [\phi_i (\phi^2)^p] = \sqrt{{2p+N \over N}} 
\left( [\phi_i] 
+  \sqrt{{18 p^2 \over N+2}} [ \phi_i \phi^2 ] \right. \cr
\!\!\!&+&  \left.
 \sqrt{ 2p^2(10p{+}N{-}6)^2 \over (N+2)(N+4) } [ \phi_i (\phi^2)^2 ] \right) 
 + \ldots \,.
\eea
Again when the interaction turns on, the operator $[ \phi_i \phi^2 ]$ becomes a descendant at $d=4k- \epsilon$, while $[ \phi_i (\phi^2)^2 ]$ becomes a descendant at $d=3k- \epsilon$. For $d=4k- \epsilon$ the matching condition is given by following two relations
\bea \label{eq:tworelations}
\frac{(-1)^{k+1}(k)_k}{4k(2k)_k}\frac{(\gamma_{p-1,1}^{(1)}-\gamma_{p,0}^{(1)})^2}{\gamma_{0,1}^{(2)}} &=& \frac{(6p+N-4)^2}{2(N+2)} \, , \cr
\frac{(-1)^{k+1}(k)_k}{4k(2k)_k}\frac{(\gamma_{p,1}^{(1)}-\gamma_{p,0}^{(1)})^2}{\gamma_{0,1}^{(2)}}
&=&\frac{18p^2}{N+2}\, .
\eea
Here we denote anomalous dimension of $(\phi^2)^{p}$ as $\gamma_{p,0} =\gamma^{(1)}_{p,0} \epsilon + \gamma^{(2)}_{p,0} \epsilon^2 + \ldots $, and similarly for $\phi_i (\phi^2)^{p}$, we have 
$\gamma_{p,1} =\gamma^{(1)}_{p,1} \epsilon + \gamma^{(2)}_{p,1} \epsilon^2 + \ldots $. 
Removing $\gamma^{(1)}_{p,0}$ from (\ref{eq:tworelations}), we obtain a recursion relation of $\gamma^{(1)}_{p,1}$, whose solution is given by \cite{comments} 
\bea
\label{ONrec}
\gamma_{p,1}^{(1)}=\beta(k,N)\, p(6p+N+2)\, , 
\eea
with $\beta^2(k,N) = (-1)^{k+1}\frac{2\,k(2k)_k}{(k)_k(N+2)}\gamma_{0,1}^{(2)}$. 
As before $p=1$ is a special case where $\phi^i \phi^2$ becomes a descendant with a fixed conformal dimension $d/2+k$, this leads to $\gamma^{(1)}_{1,1}=1$. Thus we have
\bea \label{eq:finding3}
\gamma_{0,1}^{(2)}=(-1)^{k+1}\frac{(k)_k }{2k (2k)_k}\frac{N+2}{(N+8)^2}\, .
\eea
Plugging this result back to (\ref{ONrec}) we also obtain, 
\bea \label{eq:finding4}
\gamma^{(1)}_{p,1} = { p (6p+N+2) \over N+8 }  \, , \quad 
\gamma^{(1)}_{p,0} = { p (6p+N-4) \over N+8 }  \, .
\eea
A similar analysis for the theory in $d=3k-\epsilon$ gives
\bea \label{eq:finding5}
\gamma_{0,1}^{(2)}=(-1)^{k+1}\frac{(k/2)_{k}}{8k(3k/2)_k}\frac{(N+2)(N+4)}{(3N+22)^2} \, ,
\eea
as well as
\bea \label{eq:finding6}
\gamma^{(1)}_{p,1} &=& { 10 p + 3N+2 \over 3 (22+3N)} p (2 p-1)\, , \cr 
\gamma^{(1)}_{p,0} &=& { 2(10 p + 3N-8) \over 3 (22+3N)} p (p-1) \, . 
\eea
The results of $\gamma^{(2)}_{0,1}$ for $k=1$ agree with known results using other methods~\cite{O(N)known1, O(N)known2}. By considering different correlators, one can obtain anomalous dimensions for other operators \cite{longpaper}. For instance, for the anomalous dimension $\gamma^{(1)}_{p,1,1}$ of the symmetric traceless tensor $ \phi_i \phi_j (\phi^2)^p -{1\over N} \delta_{ij} (\phi^2)^{p+1}$ we have,
\bea \label{eq:finding7}
\gamma^{(1)}_{p,1,1}  &=& \frac{2+p(8+N+6p)}{N+8}  \, ,   \quad d=4k-\epsilon \, , \\
\gamma^{(1)}_{p,1,1}  &=& \frac{2p(2+3p(N+4)+10p^2)}{3(3N+22)} \, ,  \quad d=3k-\epsilon \, . \nonumber
\eea

\section{Conclusion}
In this letter we applied general properties of CFTs,  and in particular the singularity structure of generic  conformal blocks, to study the possible smooth deformations of GFCFT's in arbitrary dimensions. Our non-trivial results correspond to Generalized Wilson-Fisher critical points.  The examples presented in this short note  include general classes of multicritical points and O$(N)$ invariant theories. Our corresponding results for theories with multiple deformations will appear in the longer version of this work \cite{longpaper}. Combining the OPE structure with
 universal properties of certain scalar null states we 
derived, at the 
first non-trivial order in the $\epsilon$-expansion, the anomalous dimensions
of an infinite class of scalar local operators. In the particular cases where 
other computational methods were applied, the results agree. Our method allows us to put huge classes of critical theories under a unified calculation scheme. We also  remark that unlike the usual conformal bootstrap program, neither crossing symmetry nor unitarity play crucial roles in our scheme. Therefore, we believe that our method, properly extended, can be useful to study non-unitary critical systems that are relevant in physics. As a final remark, we observe that we have considered even marginal deformations of the form $\phi^{2n}$. When the deformation is odd, namely  $\phi^{2n-1}$, our method appears to be less powerful. Furthermore it is of great interest to consider more general deformations (such as turning on multiple marginal interactions) and more general operators (such as operators with spins). More details on these questions as well as on the calculations presented in this letter will appear in the longer version of our work \cite{longpaper}.
\section{Acknowledgements}
\small{ Our work has been strongly inspired by some unpublished notes of Yu 
Nakayama on smooth deformations of free-field theory in $d=6-\epsilon$ 
dimensions. We thank him for sharing with us his notes. We would also like to thank M. Bianchi and D. Simmons-Duffin for helpful discussion. 
F.G. wishes to thank the OIST of Okinawa where part of this work was done. 
The work of C.W. is supported by DOE Grant No. DE-SC0010255.  A. C. P. wishes to thank CPHT \'Ecole Polytechnique, for its warm hospitality during the final stages of this work. The work of A.~L.~G.
is funded under CUniverse research promotion project by Chulalongkorn University (grant
reference CUAASC). }

\begin{thebibliography}{99}
\bibitem{comments} 
There is a sign ambiguity here, which is fixed by requiring it reproduces the results of $N=1$, namely the single scalar case we considered in the previous section.
\end{thebibliography}

\end{document}